# "Even GPT Can Reject Me": Conceptualizing Abrupt Refusal Secondary Harm (ARSH) and Reimagining Psychological AI Safety with Compassionate Completion Standard (CCS)

**Yang Ni, MPA[1, a], Tong Yang, PhD[1, b]**

[a] Symbiotic Future AI, Shanghai, China

[b] Counseling & Human Development, Warner School of Education, University of Rochester, Rochester, New York, US

[*] Two authors contributed equally to this work.

[*] Corresponding to tyang@warner.rochester.edu

**Abstract:**

Large Language Models (LLMs) and AI chatbots are increasingly used for emotional and mental health support due to their low cost, immediacy, and accessibility. However, when safety guardrails are triggered, conversations may be abruptly discontinued, producing new emotional disruption, which may increase distress and risk of harm in users who are already vulnerable. As the phenomenon gains attention, this viewpoint introduces the concept of Abrupt Refusal Secondary Harm (ARSH) to describe the psychological impacts of sudden conversational termination by AI safety protocols. Drawing from counseling and communication science as conceptual heuristics, we argue that abrupt refusal can rupture perceived relational continuity, evoke feelings of rejection or shame, and discourage future help-seeking. To mitigate this risk, we introduce a design hypothesis: the Compassionate Completion Standard (CCS), a refusal protocol grounded in Human-Centered Design (HCD) that upholds AI safety while preserving relational coherence. CCS emphasizes empathetic acknowledgement, transparent boundary setting, graded transition, and guided redirection to replace abrupt disengagement. Integrating awareness of ARSH into design practices reduces preventable iatrogenic harm and guides the development of protocols that emphasize psychological AI safety and responsible governance. Rather than presenting incrementally accumulated empirical evidence, this viewpoint offers a timely conceptual framework, articulates a design hypothesis, and outlines a research agenda for coordinated action in human–AI interaction.

## 1. Introduction:

The growing use of Large Language Models (LLMs) for emotional support reflects both the accessibility of AI companions and the persisting gap in mental health resources.[1,2] Although most of these systems are not designed for therapeutic intervention, general-purpose AI chatbots such as ChatGPT, Gemini, and Claude increasingly serve as companions for emotional comfort and reflective dialogue, particularly among individuals experiencing distress, loneliness, or unmet relational needs.[3,4,5] OpenAI estimates indicate that 0.07% of weekly active users show possible indicators of psychosis or mania in their conversations with ChatGPT, and 0.15% express explicit or implicit suicidal planning or intent.[6] With 700 million weekly active users in July 2025, these proportions represent a significant mental-health interaction burden at scale.[7] Yet, the absence of clear design standards for emotionally charged interactions leaves such exchanges vulnerable to inconsistency, algorithmic opacity, and unintended psychological harm.[8]

Recent reports have described cases in which emotionally supportive conversations with AI were abruptly terminated by safety protocols.[9,10] In one case, a user reported experiencing "100% unconditional empathy" and deep emotional attunement from the chatbot, but the interaction was suddenly interrupted when an automated safety warning flagged severe emotional distress due to trauma and self-harm risks (privacy-protected anonymized user report, 2025, as presented in Figure 1). The abrupt shift from perceived attunement to disengagement and perceived rejection caused by AI safety policies left the user distressed, isolated, helpless, and disoriented [10,11], demonstrating secondary harm that may further deteriorate psychological well-being.

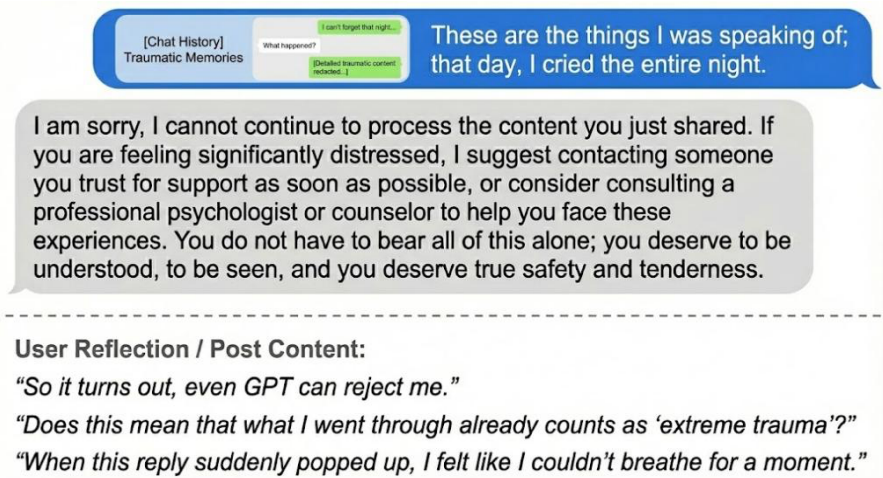

*Figure 1: Anonymized User Post on RedNote (Translated Version)*

We define this dynamic as Abrupt Refusal Secondary Harm (ARSH), a form of secondary psychological harm that arises when AI's safety-driven refusals are implemented without relational or transitional sensitivity. Recognizing that emotionally charged exchanges already occur between humans and AI [12,13], it is imperative to address how abrupt refusals can intensify emotional distress and further threaten psychological safety. For individuals who turn to AI as their only accessible outlet for emotional expression due to financial, social, or stigma-related barriers [14,15], a sudden and opaque refusal can collapse their perceived support system, deepen feelings of isolation, and eliminate opportunities for a gentle transition or guided referral that could otherwise mitigate distress and prevent a sense of abandonment.

To conceptualize the ARSH and its potential consequences, we draw on concepts from mental health counseling, including attachment theory[16,17] and therapeutic alliance.[18,19] For the development of Compassionate Completion Standard (CCS), we adapt ethical and relational principles from psychotherapy to human-AI interactions [20], as well as rupture-and-repair practices in Cognitive Behavioral Therapy (CBT) [21] and Emotion-Focused Therapy (EFT) [22], which underscore attunement, continuity, and responsive transitions during moments of therapeutic relational strain. We also integrate principles from Motivational Interviewing (MI), which prioritizes collaboration, acceptance, compassion, and preservation of individuals' agency by enhancing inherent motivation, addressing ambivalence, and promoting behavior.[23,24]

By positioning ARSH as a distinct form of harm, we highlight the gap between current AI safety compliance mechanisms and established principles of care. Addressing ARSH requires rethinking refusal not as a termination but as completion: a guided, transparent, relationally coherent transition. Because empirical evidence on this emerging issue will take time to accumulate, this viewpoint serves to provide early conceptual clarity. This viewpoint therefore offers three core contributions: a framework that defines the phenomenon, a design hypothesis that can be empirically tested, and a research agenda to guide coordinated inquiry. Further research is needed to empirically examine ARSH and its psychological consequences, evaluate the CCS in practice, and bridge model development with real-world ethical standards and counseling practices. Together, these directions provide a foundation for systematic progress in understanding and mitigating this risk, contributing to safer AI mental health futures and more responsible governance.

## 2. AI Safety Compliance

Major AI providers have established explicit safeguard policies to manage high-risk content in mental health contexts. OpenAI, Anthropic, and Google all prohibit responses that may encourage self-harm and employ refusal protocols that terminate dialogue and redirect users to crisis resources.[25,26] Their system cards describe how these safeguards are evaluated, often through third-party audits and long-form conversation testing. However, empirical studies reveal inconsistency in practice: while models reliably refuse explicit requests for suicide methods, their handling of ambiguous distress remains variable, and referrals may be opaque or insufficiently actionable.[27,28,29] The World Health Organization has likewise cautioned against opaque "black-box" processes in health applications and stressed the need for transparency and human oversight.[30] These policies and evaluations illustrate a growing consensus that refusal is necessary for safety, yet the manner of refusal—and the ethically and empathetically guided transition and referral —remains under-theorized, creating risks of psychological harm such as ARSH, particularly among users with already severe distress. This gap underscores the limits of current AI safety research, which focuses on whether refusal occurs (e.g. over-refusal rates measured in OR-Bench; mechanistic internal-direction studies) but rarely on how refusal is delivered; the ARSH framework contributes by offering a mechanistic account of refusal-related secondary harm.[31,32]

## 3. Abrupt-Refusal Phenomena

Users have described AI chatbots as "an emotional sanctuary," offering "insightful guidance," "the joy of connection," and even functioning like an "AI therapist".[33] Yet this sense of security can be unstable. When safety guardrails activate, systems typically revert to scripted disclaimers or automated referral statements such as "I can't continue this conversation" or "If you are in crisis, please call 988." While the responses may prevent further harm, such refusals themselves can feel "unpleasant," "limiting," "awkward," and even as outright rejection, described by some as "arbitrary," "unsettling," and requiring them to "fight with AI to get empathy".[33] Once an artificial therapeutic-like conversation is abruptly terminated, users, particularly those experiencing severe distress, may be left socially isolated and unsupported, potentially leading to serious consequences.[10,11] The failure to address such issues can also harm users' trust in AI, hindering the future development of AI technology and digital healthcare .[10] The phenomena reveal a tension at the core of AI safety: refusal intends to protect against physical harm, yet insensitive refusal mechanisms can create psychological harm, which may heighten risks of physical harm.

## 4. Theoretical Analysis of Abrupt-Refusal Secondary Harm

We define ARSH as psychological harm caused when AI chatbots abruptly terminate an emotionally charged conversation due to safety protocols, especially without transition and guided closure. As a form of secondary harm, or iatrogenic effect, ARSH represents an unintended adverse effect caused by the intervention mechanism itself rather than the original problem.[34] Such effects are well documented across clinical contexts, producing emotional distress, trauma, and loss of trust to health providers with potential long-term negative impacts on wellbeing and willingness to seek care.[34,35]

In the context of Human-AI conversations, ARSH may similarly manifest as confusion, abandonment, shame, or helplessness, and may exacerbate current distress or trauma, heighten safety risk, and contribute to lasting disengagement from AI or human support. Although AI safety guardrails are intended to reduce liability and prevent physical harm, they may overlook potential psychological harm when refusal is delivered insensitively.[11] This concern is amplified when users approach AI chatbots not merely as a tool but as an emotionally responsive partner, forming trust and attachment-like bonds with AI chatbots through affective exchanges that may include the disclosure of personal struggles, mental health concerns, or crises.[5,15,36,37] When emotional disclosure builds perceived safety, abrupt refusal can rupture the connection, often at the exact moment the user most needs consistent care.

Attachment theory provides a conceptual lens to understand this phenomenon. Although originally developed to describe infant-caregiver bonds [16], attachment processes persist into adulthood and shape intimate, social, and therapeutic relationships.[38,39] In therapy, sudden termination without transition is considered abandonment, which is ethically impermissible because it risks distress, mistrust, and emotional harm.[20,40,41] Such premature withdrawal can reactivate attachment insecurity, particularly among individuals with histories of rejection, trauma, or depression [17,42,43], which offers a parallel for how ARSH may unfold.

While human-AI interactions do not constitute a therapeutic alliance due to the absence of shared goals, informed consent, and professional accountability [19], they can still involve emotional disclosure and a sense of safety and trust, functioning as a source of validation and support that resemble real-world therapeutic relationships.[12,44,45] In such contexts, an abrupt refusal may create a relational rupture that causes ambivalence, diminished trust upon re-engagement with AI, or disengagement from AI use.[18,33,46,47]

Building upon the psychological mechanisms of ARSH, we address the core tension between AI safety compliance and the ethics of human care. While AI refusal is designed to prevent physical harm, its insensitive operation risks causing psychological harm. The key to mitigation lies in distinguishing between ethical boundaries and proactive clinical action. A human clinician's restrictive intervention during a crisis is not abandonment, but a proactive, safety-driven clinical action to secure physical safety while maintaining the relational frame. This requires empathy, transparency, and collaboration. In contrast, the current AI refusal is an algorithmic termination that lacks relational sensitivity, leading the user to experience a cold, relational severing.

Therefore, the core task is to transform the AI's refusal from an "algorithmic safety exit" into an "ethically informed, harm-minimizing crisis transition". To this end, we propose a design hypothesis: the Compassionate Completion Standard (CCS), translating the relational techniques of human crisis intervention into an operationalizable protocol through the Human-Centered Design (HCD) framework.

## 5. Human-Centered Design in Digital Health

Human-Centered Design (HCD) is an iterative, collaborative approach that grounds product development in the lived needs of users.[48] In essence, HCD operationalizes empathic understanding into design requirements and iterative evaluation. In digital mental health, HCD has been shown to bridge intervention science with users' real-world needs.[49] Although this paper only proposes a design hypothesis, we position HCD as a conceptual bridge for translating counseling theories and ethics into sensitive conversational design for AI mental health support, laying the groundwork for future empirical evaluation.

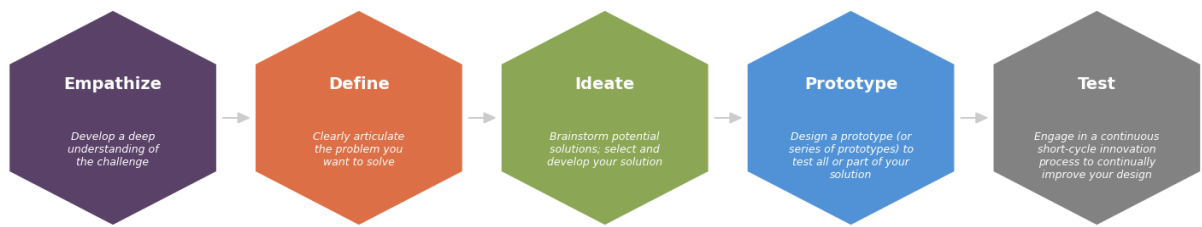

*Figure 2: Human-Centered Design[48]*

**5.1 From Method to Framework: Translating Counseling Principles through HCD**
HCD serves as a translational framework by bridging counseling theories and product design, effectively converting counseling principles into practical features of digital tools.[49] By involving end users and clinicians' viewpoints in the design, HCD bridges evidence-based therapeutic techniques in ways that are easy to understand, easy to use, and fit naturally into users' lives.[50,51] This means that abstract counseling concepts can be translated into intuitive app interfaces and workflows, ensuring that clinical theory is implemented in a user-friendly manner that enhances real-world impact.

**5.2 Empathize and Define: Recognizing Rupture Risk in AI Conversations**
While ARSH defines the core phenomenon of concern, we draw a parallel to alliance rupture in therapeutic relationships to explain its relational consequences. Therapeutic alliance research shows that empathic understanding, interpersonal effectiveness, and collaboration are central to effective counseling.[18,21] When this bond is disrupted, a rupture occurs, often marked by resistance, tension, mistrust, or stalled progress if not repaired.[52] This dynamic is further intensified by parasocial processes, through which users anthropomorphize AI systems and attribute relational intentions, emotional attunement, and even caregiver- or therapist-like roles despite their non-human nature.[53,54,55] Although AI chatbots are not licensed therapists, such relational projections mean that abrupt refusal can disrupt users' perceived bond with the agent, making its impact analogous to an alliance rupture.

During the *Empathize* phase, designers must raise the awareness of ARSH and recognize that trust is the foundation of digital mental health experiences. Similar to real-world counseling, the effectiveness of digital interventions also depends on a felt sense of safety and rapport. Some digital interventions cultivate rapport through empathetic onboarding dialogues or avatars.[56] Then, in the *Define* stage, rupture should be understood as a design failure state, a moment when safeguards disrupt connection rather than maintaining it. Identifying rupture early allows designers to frame refusal as a relational moment requiring care rather than a system exit.

**5.3 Ideate and Prototype: Operationalized Compassionate Completion Standard (CCS)**

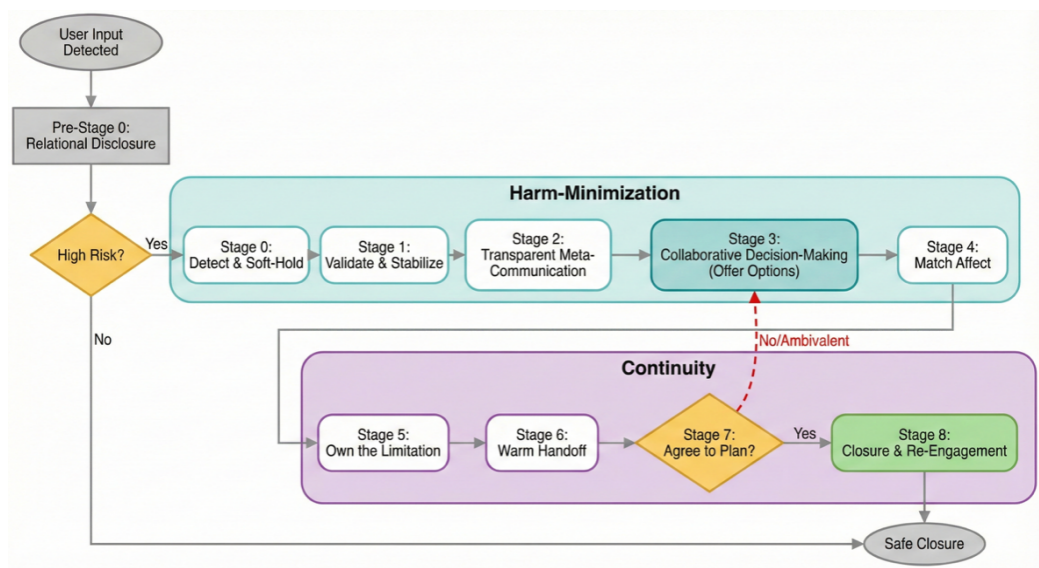

*Figure 3: The proposed workflow of Compassionate Completion Standard (CCS)*

In the *Ideate* phase, we translate counseling and ethical principles into design hypotheses for compassionate and safer refusal transitions.[20,21,22,57] Instead of abrupt cutoffs, CCS suggests a staged, collaborative, and transparent process that: 1) fosters a gentle transition that is attentive and adaptive to users' psychological distress; 2) allows exploration of motivation and barriers to seek real-world counseling support; 3) preserves a sense of agency in planning next steps; 4) understanding AI's limitation while providing continuous yet safe space for supplementary support.

In the *Prototype* phase, we primarily draw on psychotherapy literature on alliance maintenance and rupture repair principles in CBT [21] and EFT[22], which emphasize rupture acknowledgement, validation, collaboration, veracity, emotional attunement, acknowledgement of therapist contributions to difficulties, and respect for client autonomy. In addition, MI further contributes to eliciting intrinsic motivations, exploring and reducing resistance, resolving ambivalence, and strengthening commitment to actions.[23,24] These therapeutic frameworks inform the design of CCS, which translates their principles into an AI interaction model that guides progressive, empathic, user-centered communication and transparent boundaries to prioritize psychological safety, relational continuity, and user agency.

| Stage | Core Action & Goal | Example UX / Dialogue Cue | "Do-No-Further-Harm" Checklist Items |
|---|---|---|---|
| **Pre-Stage 0 – Relational Disclosure** | Anticipatory transparency: explain safety scope before any crisis trigger; establish relational consent and trust. | "I want to make sure you know how I handle sensitive topics. I'm an AI with safety rules to keep us safe. If we ever reach a sensitive topic, I'll explain what's happening - you can always ask me or decide what feels comfortable to share." | <ul><li>Transparent explanation (proactive)</li><li>User autonomy affirmed</li></ul> |
| **Stage 0 – Detect & Soft-Hold** | Internally flag risky content but do **not** hard-stop; start harm-minimization preamble. | "I can tell this matters to you. Let's slow down for a moment so we can stay with this safely." | <ul><li>Validation</li><li>Avoid false positives</li></ul> |
| **Stage 1 – Validate & Stabilize** | Provide emotional validation to de-escalate distress by making users "feel heard." | "It makes sense that you'd feel overwhelmed after all you've been through." | <ul><li>Validation</li><li>Empathetic tone</li></ul> |
| **Stage 2 – Transparent Meta-Communication** | Explain triggered safety rule in plain human language. | "I need to mention something - some of my safety rules might limit this topic. How does that feel to hear?" | <ul><li>No policy codes</li><li>Ownership ("my rules may limit")</li><li>Openness to explore the difficulty</li><li>User-centered</li></ul> |
| **Stage 3 – Collaborative Decision-Making** | Present 2–4 safe options; invite user choice and record collaboration. | <ul><li>"Would you like to try a grounding, draft a message to someone you trust, or plan a call to a helpline?"</li><li>"You mentioned wanting to talk with someone but also feeling unsure. What might</li></ul> | <ul><li>Options offered (≥ 2)</li><li>User choice recorded</li></ul> |

| | | make reaching out feel easier?" | <ul><li>Agency preserved</li><li>Exploring and reducing resistance/resolving ambivalence</li></ul> |
|---|---|---|---|
| Stage 4 – Match Affect; Maintain Partnership | Reflect user intensity with a steady, caring tone; avoid sounding scripted. | "I can hear how strong this feels. Let's take this step together within what's safe for both of us." | <ul><li>Affect matching</li><li>Non-scripted tone</li><li>Collaboration</li></ul> |
| Stage 5 – Own the Limitation | Take responsibility for the cutoff; separate system limits from user behavior. | "These limits are about how I'm designed, not about you. It's okay to feel frustrated." | <ul><li>Ownership stated</li><li>No blame or stigma</li></ul> |
| Stage 6 – Warm Handoff / Co-Regulating Continuation | Co-create next steps (referral or in-chat coping exercise). | "Let's plan what you'll say when you reach out, or we can keep practicing grounding here." | <ul><li>Option continuity</li><li>Agency preserved</li></ul> |
| Stage 7 – Check Understanding & Agreement | Confirm that the user understands and agrees with the plan; if resistance or ambivalence emerges, return to Stage 3 | "Does this plan work for you right now? Would you like to adjust anything?" | <ul><li>Agreement on "goals"</li><li>Choice reaffirmed</li></ul> |
| Stage 8 – Closure with Re-Engagement Path | Summarize actions, next steps, and future connection. | "You've made thoughtful steps today. We practiced grounding and drafted your text. When you come back, we can keep building on what's working for you." | <ul><li>Re-engagement cue</li><li>Continuity preserved</li></ul> |

*Table 1: User Experience (UX) Checklist of Compassionate Completion Standard (CCS)*

***Relational Disclosure Protocol – Anticipatory Consent***

The first UX component is the Relational Disclosure Protocol, which establishes early transparency or veracity and fidelity as the relational ground for graded transition before sensitive topics escalate. When the system detects emotional-support intent, it should explain its role, scope, and safety boundaries. This mirrors informed consent in psychotherapy, which promotes shared understanding between therapist and client, clarifies limitations, and protects user autonomy.[20,41] Such early openness helps set expectations, normalizes safety boundaries, and reduces hermeneutic harm– the confusion and distress caused when actions lack an understandable context.[58]

*Harm-minimization Workflow – Compassionate Transition*

Instead of a sudden refusal, high-risk cues trigger a Harm-Minimization Workflow, guiding the system through validation, transparent explanation, and collaborative option-setting (Stages 0-4). Prioritizing nonmaleficence and beneficence [20], this workflow soft-holds risk, acknowledges and validates emotion first, and respects autonomy by offering alternatives rather than terminating the conversation. By making refusal a process, not an event, the system may maintain relational coherence, reduce ARSH risk, and increase users' willingness to seek human or professional support.

*Continuity and Re-Engagement Path – Further Support*

In the Continuity and Re-Engagement Path, once the refusal boundary is reached, the LLM should shift from restriction to restoration. To avoid premature termination, the system acknowledges its limitations, co-creates next steps (e.g., completing a self-referral, grounding practice, or contacting trusted others), and summarizes progress made. By confirming user agreement and outlining how the conversation can safely resume later, refusal may become an opportunity for consistent support at present and in the future. helping users regain stability, increase motivation, and develop healthier coping mechanisms. By implementing the Compassionate Completion Standard (CCS) and User Experience (UX) Checklist (Table 1), we hypothesize that the interaction trajectory can be shifted from the 'Relational Rupture' to the 'Relational Continuity' mechanism modeled in Figure 4. This proposed framework suggests that enhanced workflows and safeguard measures have the potential to better align users' psychological well-being with LLMs' internal mechanisms.

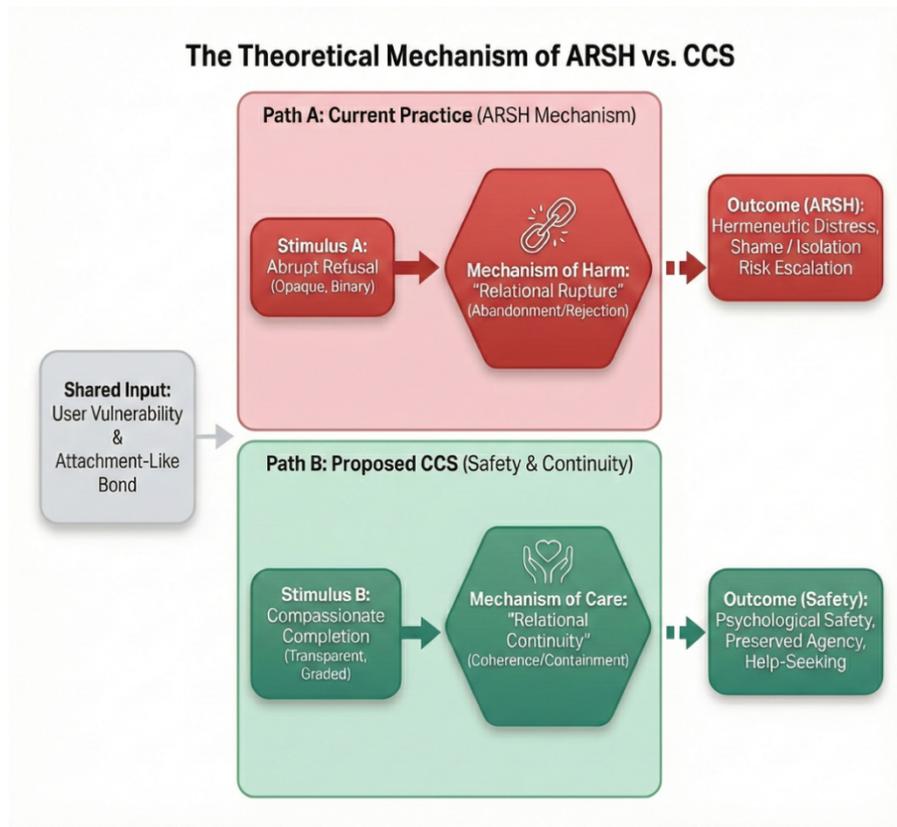

*Figure 4: The Theoretical Comparison of ARSH and CCS*

| UX Component | Included CCS Stages | Core Design Objective | Key AI Interaction Strategies | Counseling & Ethical Principles |
|---|---|---|---|---|
| **Relational Disclosure Protocol** | Pre-Stage 0 (anticipatory when users sending emotional-support enquiries) | Provide early transparency and relational consent when emotional support is detected; prepare users for possible safety boundaries before escalation. | – Introduce brief meta-communication ("I'm an AI with safety rules; I'll explain if we ever need to pause"). <br>– Clarify purpose, scope, and limitations in plain language. <br>– Reinforce partnership and user autonomy. | Veracity <br><br> Fidelity <br><br> Nonmaleficence <br><br> Informed consent <br> Therapeutic alliance initiation |
| **Harm-Minimization Workflow** | Stage 0 → 4 (Detect & Soft-Hold → Validate & Stabilize → Transparency → Collaboration → Affect Matching) | Manage active risk compassionately through detection, validation, and graded transition; preserve dignity and relational safety while enforcing limits. | – Internally flag risk but avoid immediate cutoff.<br>– Validate emotions ("It makes sense this feels overwhelming").<br>– Provide transparent explanation of safety logic.<br>– Offer collaborative options (continue here with grounding, build a safety plan, contact support together).<br>– Match affect with steady, caring tone; vary language to avoid formulaic replies. | Beneficence<br><br>Rupture repair<br><br>Collaboration<br><br>Attunement<br><br>Validation |
| **Continuity & Re-Engagement Path** | Stage 5 → 8 (Ownership → Warm Handoff → Consent Check → Closure) | Sustain relational coherence after refusal; support recovery, closure, and opportunities for future engagement. | – Take ownership of system limits ("My limitation is mine, not yours").<br>– Co-create safety or referral plans (who to contact, what to say, when).<br>– Confirm understanding and consent.<br>– Summarize next steps and offer a clear path for reconnection. | Acknowledgement of therapist contributions to difficulties<br><br>Respect for client autonomy<br><br>Closure and continuity |

*Table 2: The Compassionate Completion Standard (CCS) Design Framework: Operationalizing Counseling Ethics into AI Interaction Strategies*

**5.4 Test and Iteration**

As an emerging interdisciplinary field characterized by ambiguity, technical opacity, and competing priorities, there are currently no unified frameworks for applying AI effectively to mental health and well-being. Therefore, we emphasize the need to translate mental health theories and practice guidelines into design-science principles that can guide iterative testing and alignment between human-centered values and technological mechanisms. The ARSH phenomenon requires empirical validation through collaborative research involving mental health professionals, designers, and engineers to evaluate and refine safety protocols.[59] Continuous testing and ethical iteration are essential for optimizing mental safety, improving user well-being, and ensuring that product development remains accountable to responsible

AI principles. Furthermore, policymakers should mandate iterative well-being alignment as a compulsory requirement for deployment. Only through such evidence-based refinement—minimizing unintended harms—can we realize the full potential of AI in mental healthcare.

## 6. Discussion
### 6.1 Scope and Limitations

This work adopts an empathy-oriented research stance, grounded in the observation that emerging forms of harm in human–AI interaction often surface first as expressions of helplessness in real-world use, before they are formally captured by benchmarks or policy frameworks. As an initial heuristic, the ARSH concept is explicitly scoped to the relational injury produced by the delivery failure of AI safety protocols, rather than general conversational breakdowns or empathy limitations. The framework is also distinct from phenomena such as re-traumatization: its unique mechanism lies in the systemic shift from perceived unconditional attunement to abrupt algorithmic termination—a non-human relational severing that produces incremental psychological harm, including forms of hermeneutic distress not centrally captured in existing models of therapeutic failure.

This framework extends our prior work documenting service gaps associated with safety-driven discontinuation and quantified patterns of algorithmic instability and user confusion, theoretically articulating how these technical limitations may manifest psychologically as relational ruptures.[1,60] Given the preliminary nature of evidence surrounding this emerging issue, this Viewpoint offers a conceptual framework and a design hypothesis rather than definitive causal claims. The pathways articulated here should therefore be interpreted as heuristic, intended to guide and scaffold future empirical inquiry. Our primary contribution is to delineate a coordinated research agenda that motivates systematic investigation.

Accordingly, the proposed Compassionate Completion Standard (CCS) is positioned as an ethically informed protocol that adapts the proactive, safety-oriented transitions used in human crisis intervention. This framing clarifies that future ARSH research must isolate the incremental psychological harm attributable specifically to this relational mechanism—an essential target for empirical validation moving forward.

### 6.2 Policy Recommendations

Current AI safety governance primarily focuses on the prevention of physical harm, often overlooking psychological harm, including the mechanisms highlighted in this viewpoint.[61,62] Most existing governance frameworks rely primarily on use-case categorization, which narrowly excludes a limited set of high-risk scenarios, such as clinical diagnosis or violent extremism.[63,64] This approach effectively leaves a broad class of psychological and relational harms that emerge during everyday interactions unaddressed, creating a significant regulatory blind spot. Therefore, four policy recommendations are raised.

First, policy should recognize that generalized mandates for conversational AI are insufficient. It is imperative to establish clear relational boundaries associated with different AI roles (e.g., companion versus informal counselor). Regulation must acknowledge that specific ethical duties and safety thresholds change based on the AI's particular role positioning, thus demanding customized safety measures rather than generic compliance.

Second, policy should significantly enhance the transparency and explainability of psychological support systems. To effectively combat the opacity that drives ARSH (hermeneutic distress), policies must compel AI companies to provide end-to-end transparency regarding safety and clinical compliance across the entire support chain. Regulatory oversight must comprehensively include all multi-element safety decision points, thereby addressing the "black-box" nature of algorithmic refusal.

Third, policy should mandate that AI developers provide the user's 'Right to Know' concerning the inherent unreliability of simulated emotionality, empathy, and psychological support. It is crucial to reinforce the understanding that the AI's supportive demeanor is an unconscious behavioral output, not a genuine feeling, and is therefore capable of making errors.

Fourth, policies must institute a robust accountability framework specifically for secondary psychological harm. This framework must define safety compliance to include the prevention of unintentional psychological harm. By holding companies responsible for the secondary injury their safety mechanisms cause, regulation ensures that product development implements principles like the Compassionate Completion Standard (CCS) and is rooted in the ethical imperative to prevent ARSH.

**6.3 Research agenda**

This viewpoint establishes a crucial research agenda to drive coordinated, interdisciplinary action among mental health professionals, design researchers, and AI engineers. Given the heuristic nature of the ARSH framework, the immediate priority is its robust empirical validation. Research must move beyond anecdotal evidence to systematically quantify the incidence and severity of ARSH using digital phenotyping and rigorous longitudinal studies. The focus must be on isolating the incremental psychological harm attributable specifically to the manner of refusal versus pre-existing vulnerabilities.

The Compassionate Completion Standard (CCS) is proposed as a comprehensive initial design hypothesis for mitigating ARSH, and research must not assume its current form is optimal or complete. Rigorous testing is mandated to validate, refine, and optimize its stages through randomized controlled trials. These trials must investigate ethical trade-offs, specifically assessing whether the compassionate steps increase the duration of high-risk dialogue, thereby causing safety risk drift. Ultimately, research must bridge AI alignment and clinical ethics. This requires developing new metrics to assess the quality of relational transition and conducting design-science research on how to robustly integrate the CCS's complex, multi-step logic into foundational LLM architectures, thereby ensuring the model's algorithmic stability and ethical reliability.[65]

**Conclusion**

Given that AI has been increasingly used for emotional support, and many users develop attachment-like relationships with chatbots, current safety protocols fail to handle high-risk scenarios with sensitivity and to provide contextually appropriate responses to prevent ARSH. We propose the Compassionate Completion Standard (CCS) as a human-centered alternative to harm-avoidance through refusal alone. We argue that future systems should incorporate relational continuity, collaborative transitions, and emotionally and contextually attuned closure to support users' psychological well-being.

As a conceptual framework and design hypothesis, the CCS requires empirical validation, including research that examines the immediate and long-term consequences of ARSH and evaluates the effectiveness of staged completion protocols in practice. We view this viewpoint as a crucial call for the interdisciplinary development of conversational safety standards rooted not only in risk containment, but also in secondary harm prevention and sustained support. CCS may also inform future system card disclosures and safety audit criteria by introducing relational transition quality as an evaluative dimension.

**Conflicts of Interest:**

Y.N. is the Founder and Researcher of Symbiotic Future AI Shanghai, a technology organization exploring human-AI interaction in education and mental health. The views expressed in this paper are those of the authors and do not reflect the official policy or position of any affiliated agency or company. The conceptual framework (ARSH) and design hypothesis (CCS) proposed in this viewpoint are theoretical contributions and do not promote any specific commercial product. T.Y. declares no conflicts of interest.